\documentstyle[pre,aps,twocolumn,epsfig]{revtex}
\begin{document}
\draft \flushbottom

%-----------------> Title etc. 
%
%
\title{Correlation Structure and Fat Tails in Finance: a New Mechanism.}
\author{Marco Airoldi}
\address{
	Risk Management \& Research, Intesa--Bci Bank \\
        Via Clerici 4, Milan, ITALY 
	% \\
        % E--mail: MarcoAiroldi@bancaintesa.it
	} 

\date{\today}

\maketitle

%-----------------> Abstract
%
%
\begin{abstract}

Fat tails in financial time series and increase of stocks cross-correlations 
in high volatility periods are puzzling facts that ask for new paradigms. 
Both points are of key importance in fundamental research as well as in Risk 
Management (where extreme losses play a key role). \\ 
In this paper we present a new model for an ensemble of stocks that aims to 
encompass in a unitary picture both these features. Equities are modelled 
as quasi random walk variables, where the non-Brownian components of stocks 
movements are leaded by the market trend, according to typical trader 
strategies. \\
Our model suggests that collective effects may play a very important role 
in the characterization of some significantly statistical properties of 
financial time series. 
 
\end{abstract}



% Pacs code description:
%
% 05.40.-a Fluctuation phenomena, random processes, noise, and Brownian motion
% 	05.40.Fb Random walks and Levy flights
% 	05.40.Jc Brownian motion
%	
% 89.65.-s Social systems
%	89.65.Gh Economics, business, and financial markets
%
% 89.75.-k Complex systems
%	89.75.Da Systems obeying scaling laws
%
% 01.75.+m Science and society

% \maketitle

%-----------------> Section: Introduction
%
%
\section{Introduction} \label{section:introduction}

One of the most important goals in the application of Physics to 
Finance~\cite{Mantegna_0,Stanley} and in Risk Management, is to model 
accurately financial time series. \\ 
Indeed physicists have become more involved in the study of financial systems 
and financial time series, among all complex systems, due to the huge amount 
of available data in electronic format, a situation that makes easier to 
check paradigms and theories. \\
On the other hand, practitioners working in finance need to understand 
financial time series in order to: (i) assess correctly the market risk
associated to portfolios and (ii) make a more rational price evaluation of 
derivative products, a task where the knowledge of stochastic process 
followed by the underlying asset is a key element. (E.g. the largely 
accepted Black--Scholes formula for option pricing~\cite{Black} is based on 
the Gaussian hypothesis for asset returns. 
Moreover many attempts to justify smiles in implied volatility structures 
are based to alternative dynamics for price variations~\cite{Smile}.) \\

The modelling of financial time series started at the beginning of the 20th 
century. In a pioneering work, Bachelier showed that stock price fluctuations 
could be modelled as a random walk (i.e. a Brownian motion)~\cite{Bachelier}. 
Indeed more accurate analyses of financial data have shown that 
geometric Brownian motion is inadequate to describe correctly 
empirical distributions when large events 
occur~\cite{Mandelbrot_1,Mantegna_1,Mantegna_2,Mantegna_3,Meyer,Ghashghaie}. 
In order to account for fat tails in empirical data, Mandelbrot, in his 1963 
paper, proposed to model financial time series through a levy stable 
process~\cite{Mandelbrot_1}.
% ~\cite{Levy,Mandelbrot_1}. 
More recently Mantegna and Stanley pointed out that a levy flight distribution 
is an accurate description only for the central part of the distribution of 
returns while for the extreme tails an exponential fitting is more appropriated; 
in particular they suggested that the distribution of price changes is 
consistent with a truncated L\'evy flight distribution~\cite{Mantegna_1}. \\
Other empirical investigations considered the time evolution of 
volatility (i.e. the second moment of price returns 
distribution)~\cite{Mantegna_2,Dacorogna_1,Scalas,Mantegna_0}. 
The analysis of high frequency data has shown that two different regimes take 
place: a super-diffusive behaviour for short time intervals (where volatility 
scales with time as $\sigma(t) \sim t^{\beta}$, $\beta \approx 0.8 > 1/2$) 
and a regime close to diffusive behaviour ($\sigma(t) \sim t^{1/2}$) for 
long times (see~\cite{Mantegna_0}, par.~7.1). Clearly the standard geometric 
Brownian motion does indeed account only for a diffusive regime. \\    
The discussions made above concern the univariate case, where only a single 
financial time series is considered. The extension of the one-dimensional 
Brownian motion to a set of $N$ assets is the so-called multivariate 
geometric Brownian motion. Today most of the analyses in Risk Management are 
based on this model, in spite of its limitations~\cite{Embrechts_1}. Within 
the multivariate Brownian motion the cross-correlations between stocks are 
supposed to be constant and independent from market activity. Indeed 
empirical evidence has shown that correlations between stocks are strictly 
dependent from market volatility and increase sharply during 
turmoil~\cite{Lin,Erb,Solnik,Longin_2,Ang,Drozdz,Bouchaud}, a phenomenon 
called correlation breakdown.  Moreover it has been demonstrated that a couple 
of correlated series, following a geometric Brownian motion, are asymptotically 
independent (i.e. extreme events occur independently)~\cite{Sibuya,Embrechts_1}. 
This means that the covariate model is unable to describe the correlation 
structure of extreme returns. All these facts have serious consequences on 
Risk Management~\cite{Markowitz,Elton}. In fact, if we use a covariate model 
to estimate the market risk (i.e. the risk due to unexpected variations of 
market conditions) of a portfolio of stocks, we under-estimate the probability 
of simultaneous losses and consequently under-estimate the possible portfolio 
claims. On the other hand, from a theoretical point of view, in recent years 
an increasing attention has been devoted to the study of ensembles of stocks 
traded in the same financial market~\cite{Lillo}, with an emphasis on 
correlation structure characterization~\cite{Bouchaud}.

Starting from this context, we introduce a new model for a set of stocks that 
aims to encompass in a unitary picture some of the most important features of 
financial time series, that is: time evolution of volatility, fat tails and 
correlation structure. For this model we show that a super-diffusive process 
characterizes the time scaling of volatility for intermediate time intervals 
(with a critical exponent 3/4) while a standard diffusive regime takes place 
for long times (par.~\ref{subsection:power_law}). 
In paragraph~\ref{subsection:fat_tails} we discuss how fat tails can emerge 
in the system. In paragraph~\ref{subsection:correlation} a Monte Carlo 
simulation approach is used to investigate the correlation structure of the 
model. Finally Section~\ref{section:conclusions} is devoted to discuss briefly 
some conclusions.

%-----------------> Section: The model
%
\section{The model} \label{section:the_model}

As discussed in the Introduction, one of the most interesting characteristics 
of financial markets is the increase of cross-correlations when large market 
events occur. Empirical evidence shows that for large market returns almost 
90\% of the equities have the same sign as that of the market~\cite{Bouchaud}. 
In other words during market crisis or running days most of stocks follow 
the general trend. This suggests that the probability for an individual stock 
to have a positive (negative) return can be affected by a market increase 
(decrease). Basing on this consideration we introduce a one step model for 
an ensemble of $N$ equities, $\{S_i\}_{i=1}^{N}$, where each stock movement, 
$\delta S_i=\pm s$, follows a quasi random walk, with a hopping probability, 
$P_{\delta S_i}$, depending on the previous market return. Precisely: 
\begin{equation}
P_{\delta S_i^{(t)}} \big( M^{(t-\Delta t)} \big) = \frac{1}{2} + 
\frac{1}{2} \, \frac{\delta S_i^{(t)}}{s} \, 
g\big( M^{(t-\Delta t)} \big) \; ,
% \left( M - \gamma M^{3} \right) \; , 0 < \gamma < 1 \; , 
\label{eq:basic_equation_P}
\end{equation}
where:
\begin{equation}
M^{(t-\Delta t)} = \frac{1}{N} \, \sum_{i=1}^{N} 
\, \frac{ \delta S_i^{(t-\Delta t)} }{s} \;, 
\label{eq:M_definition}
\end{equation}
is the normalized market return, $\Delta t$ is the time required to complete 
one step and $g$ is a function such that $\mid g(M)\mid \le 1$. (To simplify 
the formulas in the following we have chosen $s=1$ and $\Delta t=1$.)  \\
The factor 1/2, in equation~(\ref{eq:basic_equation_P}), represents the usual 
random walk contribution (i.e. a 50\% chance to perform a positive or a 
negative step). The function $g$ incorporates the typical trader strategy: when 
market blows up some traders (the bulls) assume that the market will continue 
to rise (this behaviour is modelled by imposing the following condition on 
$g$: $\lim_{M \rightarrow 0} \frac{dg}{dM} = 1$).
On the opposite hand other agents will consider this positive trend a good 
opportunity to realise a profit by selling their positions (thus we require 
that $\mid g(M) \mid < \mid M \mid$ for $M \ne 0$). 
Finally, a rising market reflects into an upward general tendency for the 
stocks belonging to that market (as in crash periods most of equities drop 
dramatically as the market does); hence $g$ must have the same sign of $M$. \\
Since for large $N$ the normalized market return, $M$, is expected to be small 
compare to 1, it turns out that: 
\begin{itemize}
	\item[(i)] 
	the function $g$, satisfying all the above constraints, can be well 
	approximated as:
	\begin{equation}
	g(M) = M - \gamma M^{3} \; \; \; \; \; \; \; 0 < \gamma \le 1  \; ; 
	\label{eq:function_g}
	\end{equation}
	\item[(ii)]  
	$g$, in eq.~(\ref{eq:basic_equation_P}), is indeed a correction 
	respect to the random walk contribution. 
\end{itemize}

Some considerations can be made on the model defined above. First of all we 
observe that the model (with the choice~(\ref{eq:function_g})) contains a 
symmetry between upward and downward movements. Actually, as pointed out 
in the literature (see for instance~\cite{Bouchaud}), the increase of 
correlations for large market events is thought to be asymmetric, with a 
stronger interlink dependence between equities during crash periods. Indeed, 
it is very simple to incorporate this feature in our model by considering two 
different values of $\gamma$: a $\gamma^+$ for upward moves and a lower value, 
$\gamma^-$, for downward moves. However in order to simplify the analysis we 
maintain $\gamma^+=\gamma^-$. \\
Secondly, from a heuristic point of view the model proposed has a strong 
analogy with a critical Ising model in infinite dimensions, it is therefore 
not surprising that critical (i.e. power law) behaviours can emerge.  \\
Finally, it is clear that the model we have introduced is quite schematic: we 
treat all the equities on the same foot and apart from correlations induced 
by the market growth (see par.~\ref{subsection:correlation}), no explicit 
correlations are considered. However we can expect that a more complete model 
would maintain the basic features we are going to discuss.

%-----------------> Section: Results and Discussion
%
%
\section{Results and Discussion} \label{section:results}

In this section we present, relatively to our model, some results regarding 
fat tails and correlation structures. We resort to analytical calculations 
(paragraphs~\ref{subsection:market_return} and~\ref{subsection:power_law})
as well as Monte Carlo simulations (par.~\ref{subsection:correlation}).

%------> Subsection: Probability distribution of market return
%
\subsection{Probability distribution of market returns} \label{subsection:market_return}

\begin{figure}[hbp]
% Command to be used with package graphicx
\epsfig{file=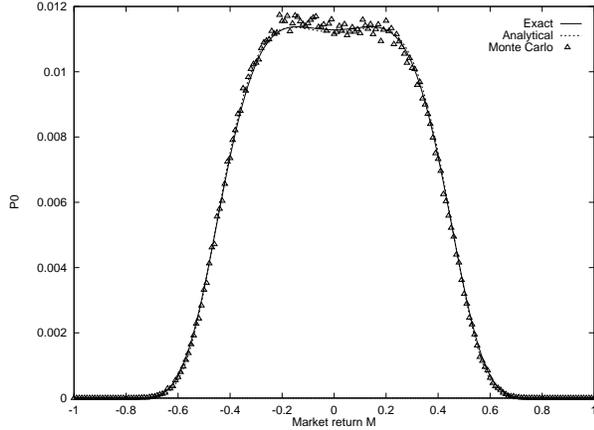,scale=0.32,angle=-90}
\caption{
The distribution of market returns over a time horizon $\Delta t$ (i.e. one 
step) for $N=200$ and $\gamma=0.2$. Solid line: exact result; dotted line: 
analytical expression~(\ref{eq:P0_analytical}); points: Monte Carlo simulation. 
}
\label{Fig:market_return_distribution}
\end{figure}

The model discussed in the previous section, can be regarded as a system of 
$N$ random walkers with a hopping probability depending on the previous mean
hopping. \\
If we indicate with $P(j,t,M)$ the probability to find a generic walker at 
site $j$, after a time $t$, when the mean hopping at time $t-1$ is $M$, then 
the time evolution master equation for $P$ is given by:
\begin{eqnarray}
&& P(j,t+1,\tilde M) = \sum_{M=-1}^{1} G(M,\tilde M) \cdot \nonumber \\
&& \cdot \left[ \frac{1+\tilde M}{2} P(j-1,t,M) + 
\frac{1-\tilde M}{2} P(j+1,t,M) \right] \, , 
\label{eq:master_prob_equation}
\end{eqnarray}
where:
\begin{eqnarray}
&& G(M,\tilde M) = \frac{N!}{ \left( N \cdot \frac{1+\tilde M}{2} \right)!
\left( N \cdot \frac{1-\tilde M}{2} \right)! } \cdot \nonumber \\ 
&& \cdot \big[ P_{\delta S=s}(M) \big]^{ N \frac{1+\tilde M}{2} }  \,
\big[ P_{\delta S=-s}(M) \big]^{ N \frac{1-\tilde M}{2} } \;,
\label{eq:G_definition}
\end{eqnarray}
and $P(j,t,M)$ obeying to the initial condition: \\
$P(j,t=0,M) = \delta_{j,0} \cdot \delta_{M,0} $. \\
Interestingly, $G(M,\tilde M)$ can be regarded as a sort of transition matrix 
from a state characterized by a market return $M$ to a state $\tilde M$.  \\
The probability distribution of market returns over a time horizon of one step
is given by $P_0(t,M) = \sum_{j} P(j,t,M)$. 
From eq.~(\ref{eq:master_prob_equation}), it follows that:
\begin{equation}
P_0(t+1,\tilde M) = \sum_{M=-1}^{1} G(M,\tilde M) \, P_0(t,M) \;. 
\label{eq:P0}
\end{equation}
After a transient period, the solution of eq.~(\ref{eq:P0}) converges to a 
stationary state. 
In fig.~\ref{Fig:market_return_distribution}, the solution obtained by solving 
numerically the eq.~(\ref{eq:P0}) is presented (solid line)  \\
%
%
%
%----> Fig. 1
%
%
%----> End fig. 1
%
%
%
An analytical solution of eq.~(\ref{eq:P0}) can be worked out, in the 
limit of large $N$, by approximating the operator 
$\hat T( f(\tilde M) ) \stackrel{def}{=} \sum_{M} G(M,\tilde M) f(M) $ 
through differential operators. Indeed by using the Stirling formula and 
considering the Taylor expunction of $f$ at the second order in 
$\delta M=M-\tilde M$, it is possible to show that: 
\begin{eqnarray}
\sum_{M=-1}^{1} G(M,\tilde M) \, && f(M) \, \approx \, f(\tilde M) + 
\nonumber \\
+ \frac{ \partial}{\partial \tilde M}  
\left[ \gamma \tilde M^3 f(\tilde M) \right] + 
\frac{1}{2N} && \frac{\partial^2}{\partial \tilde M^2} 
\left[ \left(1-\tilde M^2\right) f(\tilde M) \right] \;, 
\label{eq:G_approx}
\end{eqnarray}
where the above approximation holds in the limit of large $N$ and under the 
following restrictions on the generic function $f$:
\begin{mathletters}
\begin{equation}
\gamma M^2 \ll 1 \; \; \; \;  \forall M: f(M) \text{ is significantly} \neq 0  \; , 
\label{eq:G_approx_validity_1}
\end{equation}
\begin{equation}
\frac{1}{f(M)} \frac{d^2f(M)}{dM^2} \ll N \;. 
\label{eq:G_approx_validity_2}
\end{equation}
\label{eq:G_approx_validity}
\end{mathletters}
Setting $f=P_0$ and substituting eq.~(\ref{eq:G_approx}) in eq.~(\ref{eq:P0}), 
we obtain a second order differential equation for $P_0$; its analytical 
solution is given by: 
\begin{eqnarray}
P_0(M) & = & C \, \left( 1 - M^2 \right)^{ \gamma \, N - 1 } \,
e^{ \gamma \, N M^2} \approx  \nonumber \\ 
& \approx & C \, e^{ - \frac{\gamma \, N}{2} M^4 + M^2 }  \;,
\label{eq:P0_analytical}
\end{eqnarray}
where $C$ denotes a constant such that $\sum_{M} P_0(M) = 1 $. Interestingly, 
we observe that the above solution satisfies, for $N \rightarrow \infty$, 
both conditions~(\ref{eq:G_approx_validity}). As a definitive check we compare, 
in figure~\ref{Fig:market_return_distribution}, the probability distribution 
of market returns obtained with different methodologies, that is: 
(i) by solving numerically the exact equation~(\ref{eq:P0}); 
(ii) by using the analytical expression~(\ref{eq:P0_analytical}) and 
(iii) by resorting to a Monte Carlo simulation. 
All the three curves are quite close (the analytical and the exact solutions
are almost indistinguishable). 

It is interesting to compare the analytical solution~(\ref{eq:P0_analytical}) 
with the corresponding expression for a set of $N$ independent equities. 
In such a case the market return is the sum of $N$ uncorrelated variables 
assuming the values $\pm s$. By applying the central limit theorem, the 
probability distribution of market variations turns out to be the usual 
Gaussian function:
\begin{equation}
P_0^{\text{un-corr.}}(M) = \frac{1}{ \sqrt{2 \pi N} } \, 
e^{ -\frac{M^2 N}{2} } \;. 
\label{eq:P0_independent_stocks}
\end{equation}
As one can realise immediately, the market volatility, $\sigma_M$, computed 
from the above probability distribution, scales with $N$ as $N^{-\frac{1}{2}}$, 
in contrast with our model where $\sigma_M \sim N^{-\frac{1}{4}}$. This means 
that the inter-dependence among equities has the effect to maintain the market 
volatility still very large even for large value of $N$. As a consequence the 
extreme market price movements in our model happen far more often than would 
be expected by chance. This characteristic is one of the statistical 
properties of real financial indexes pointed out in literature~\cite{Bouchaud}.

%------> Subsection: Power law in time scaling of volatility
%
\subsection{Power law in time scaling of volatility} \label{subsection:power_law}

\begin{figure}[hbtp]
% Command to be used with package graphicx
\epsfig{file=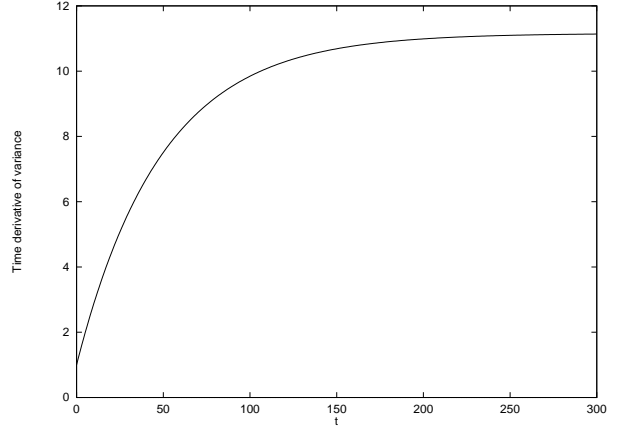,scale=0.32,angle=-90}
\caption{
Time derivative of variance, 
$\frac{d\sigma_{S}^2(t)}{dt}$, versus time, for $N=200$ and $\gamma=0.1$.
For large $t$ (i.e. $t$ comparable to $N$), the curve converges to a 
constant value (standard diffusive regime). Indeed for $N \sim 100$ the 
range of scaling behaviour is very narrow.
}
\label{Fig:volatility_time_evol}
\end{figure}

The volatility of a single stock over a horizon $t$, $\sigma_{S}(t)$, 
(which, in our model, is indeed the same for all equities) is given by: 
$\sigma_{S}^2(t) = \sum_{j,M} j^2 \, P(j,t,M)$. \\
By using the eq.~(\ref{eq:master_prob_equation}), it turns out that 
$\sigma_{S}(t)$ satisfies the following finite-difference equation:
\begin{mathletters}
\begin{equation}
\sigma_{S}^2(t+1) - \sigma_{S}^2(t)  =  
1 + 2 \cdot \sum_{M=-1}^{1} g(M) \cdot P_1(t,M)    \;,
\label{eq:volatility_diff_eq_1}
\end{equation}
where $P_1(t,M) \stackrel{def}{=} \sum_{j} j \, P(j,t,M)$ must obey to:
\begin{equation}
P_1(t+1,\tilde M) = \sum_{M=-1}^{1} G(M,\tilde M) P_1(t,M) + 
\tilde M  \, P_0(\tilde M) \;, 
\label{eq:volatility_diff_eq_2}
\end{equation}
with the initial condition $P_1(t=0,M)=0$.  \\
\label{eq:volatility_diff_eq}
\end{mathletters}
Note that if we substitute to the matrix $G(M,\tilde M)$ its approximated 
expression~(\ref{eq:G_approx}) and represent the finite difference 
$P_1(t+1,\tilde M)-P_1(t,\tilde M)$ by a time derivative, the 
equation~(\ref{eq:volatility_diff_eq_2}) reduces to a Fokker--Planck 
equation with a drift $-\gamma M^3$ and a diffusion coefficient 
$\frac{1-M^2}{2N}$ (the factor $M P_0$ can be always re-adsorbed by redefining 
$\hat P_1(t,M) = P_1(t,M) - p(M)$, for an appropriate function $p$).  \\
Regarding the time evolution equation of 
volatility~(\ref{eq:volatility_diff_eq_1}), the first contribution represents 
the usual diffusive process while the second term is originated by the 
influence of market trend on stock returns. If this last term has a power law 
time dependence $t^{\delta}$ ($\delta > 0$) then the stock volatility is 
characterized by a super-diffusive process. \\
In order to investigate the time scaling behaviour of 
equations~(\ref{eq:volatility_diff_eq}), we set $P_1(t,M) = h(t,M) \cdot P_0(M)$. 
Substituting~(\ref{eq:G_approx}) into~(\ref{eq:volatility_diff_eq_2}), the 
equation for $h$ reads: 
\begin{equation}
\frac{\partial h} {\partial t} = -\gamma M^3 \frac{\partial h}{\partial M} + 
\frac{1}{2 N} \left( 1-M^2\right) \frac{\partial^2 h}{\partial M^2} + M \;. 
\label{eq:PDE_h}
\end{equation}
Interestingly, we note that the analysis of the finite-difference version of 
eq.~(\ref{eq:PDE_h}) shows that $P_1$ can be always expressed as a finite sum 
of the form $P_1(t,M) = \sum_{k=1}^{t} a_{k,t} \, M^{2k-1} \, P_0(M)$. 
Now it is simple to prove that $M^n  \, P_0(M)$ fulfil the 
conditions~(\ref{eq:G_approx_validity}) only for 
$\sqrt { n \, \gamma / N} \ll 1$ and therefore we can expect that $P_1$ 
satisfies the relations~(\ref{eq:G_approx_validity}) only in the range of 
times $t$ such that $\sqrt { t \, \gamma / N} \ll 1$.
Hence the Partial Differential Equation (PDE)~(\ref{eq:PDE_h}) can be regarded 
as a good approximation for the evolution of $h$ only for $t \ll N/\gamma$. 
As a result the time evolution of $\sigma_{S}(t)$ is characterized by three 
regimes: 
\begin{itemize}
	\item[(i)]  
        $\sqrt{ \frac{N}{\gamma} } \ll t \ll \frac{N}{\gamma}$. $h(t,M)$ 
	satisfies the PDE~(\ref{eq:PDE_h}). The long time scaling behaviour 
	of $h$ is investigated in Appendix~\ref{section:appendix}. Precisely, 
	it can be proved that if $t \gg \sqrt{ \frac{N}{\gamma} }$ then 
	$\sum_M g(M) \, P_1(t,M) \sim t^{\frac{1}{2}}$  and therefore 
	$\sigma_{S}(t) \sim t^{\frac{3}{4}}$. In such a case the system is 
	characterized by a super-diffusive process. Note that the critical 
	exponent 3/4 is quite close to the empirical result of 0.8 reported
	in literature (see~\cite{Mantegna_0}, par.~7.1). 
	\item[(ii)] 
        $t$ comparable to $\frac{N}{\gamma}$. The PDE~(\ref{eq:PDE_h}) fails 
	in describing the time evolution of $h$. The numerical solution of
	equation~(\ref{eq:volatility_diff_eq_2}) shows that $P_1(t,M)$ converges 
	towards a stationary state (see fig.~\ref{Fig:volatility_time_evol}). 
	As a consequence $\sigma_{S}(t) \sim t^{\frac{1}{2}} $ and therefore 
	the super-diffusive scaling behaviour breaks down;
	\item[(iii)]
	an intermediate regime between the two, where volatility scales faster 
	than $t^{\frac{1}{2}}$ but without a sharp power law behaviour.
\end{itemize}
%
%
%
%----> Fig. 2
%
%
%----> End fig. 2
%
%
%
The behaviour reported above for the time evolution of volatility is 
indeed typical of financial time series (see~\cite{Mantegna_0}, par.~7.1).

%------> Subsection: Fat tails
%
\subsection{Fat tails} \label{subsection:fat_tails}

Using the above results we can have some insights on the presence of fat tails 
in the probability distribution of stock price returns. Indeed, we can expect 
that for $N$ sufficiently large, most of the time the market growth, $M$, 
is relatively small (from eq.~(\ref{eq:P0_analytical}) 
$lim_{N \rightarrow \infty} P_0(M) = 0$) and therefore the majority of stocks 
follow a geometric Brownian motion ($P_{\delta S_i}(M) \approx 1/2$). On the 
other hand during extreme market movements (crashes or running days), the 
walkers have a higher probability, respect to a simple random walk, to move 
far from the origin (extreme events). 
Therefore is reasonable to imagine that the central part of the stock price 
variations distribution is approximately near to a standard Gaussian
distribution while for the tails a fatter non-Gaussian shape is expected. 
Indeed the results of subsection~\ref{subsection:power_law} show that the 
stock volatility grows faster than the Brownian term $\sqrt{t}$ (at least 
until $t$ is of order $N/\gamma$). This is coherent with the hypothesis of 
fat tails in the distribution of stock returns, which indeed provide an 
extra-contribution to the volatility growth. Since the time evolution of 
volatility reaches a standard diffusive behaviour for $t$ comparable to 
$N/\gamma$ (as indicated by the numerical simulation reported in 
fig.~\ref{Fig:volatility_time_evol}) we can suppose that also fat tails 
would disappear on long time scales, reducing the process to a pure Gaussian 
stochastic random walk. This is in accordance with the empirical evidence 
observed in financial time series, where a crossover between a leptokurtic 
distribution for short times and a Gaussian behaviour for long time intervals 
takes place\cite{Mantegna_0}.

%------> Subsection: Cross-correlations during turmoil
%
\subsection{Cross-correlations during turmoil} \label{subsection:correlation}

\begin{figure}[htbp]
% Command to be used with package graphicx
\epsfig{file=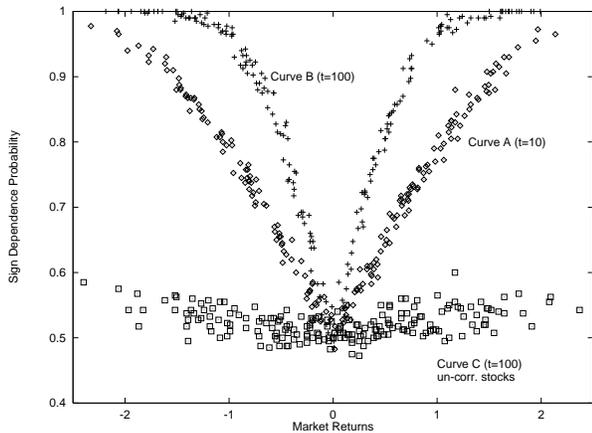,scale=0.32,angle=-90}
\caption{
Conditional probability for a stock to have the same sign of the market 
return for different values of the time horizon: $t=10; 100$ (curve A and B). 
In the figure is also reported the probability distribution corresponding to 
a set of independent stocks over a time horizon $t=100$ (curve C). 
The Monte Carlo results are obtained considering $200$ samples, $N=200$ and 
$\gamma = 0.2$.  On the x-axis the market returns (over a time $t$) are 
measured in unit of market standard deviation (over a horizon $t$). 
}
\label{Fig:cond_probab_sign}
\end{figure}

\begin{figure}[htp]
% Command to be used with package graphicx
\epsfig{file=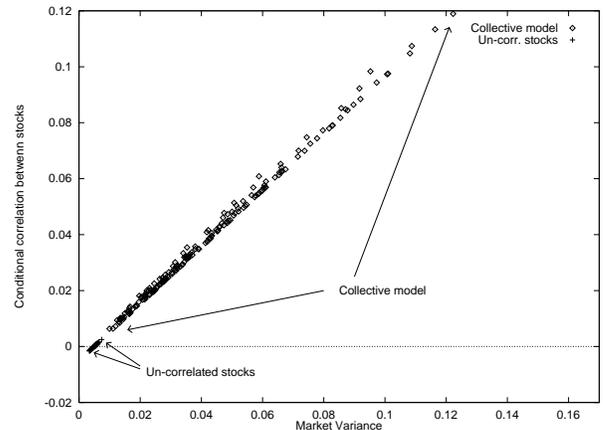,scale=0.32,angle=-90}
\caption{
Conditional correlation between stocks as a function of market 
variance $\sigma_M^2$. As the market becomes more volatile the stocks become 
more interlinked. The results in the figure are obtained using a Monte Carlo 
with $200$ samples, an observation interval of $100$ steps, $N=200$ and 
$\gamma = 0.2$.
}
\label{Fig:corr}
\end{figure}

As we have seen in the introduction, ensembles of stocks are characterized by 
an increase of correlations during critical periods. This behaviour is 
qualitatively well reproduced within our model.  \\
In figures~\ref{Fig:cond_probab_sign} and~\ref{Fig:corr}, some Monte Carlo 
results are given. \\
%
%
%
%----> Fig. 3
%
%
%----> End fig. 3
%
%
%
Specifically, in fig.~\ref{Fig:cond_probab_sign}, we present the conditional
probability for stocks to have the same sign as that of the market. The 
results show a strong sign dependence as a function of the market return. 
Interestingly, that dependence increases further considering larger time 
horizons (compare curve A and B in fig. ~\ref{Fig:cond_probab_sign}). 
On the other hand, for a set of uncorrelated stocks, the sign conditional 
probability is substantially weaker (see fig.~\ref{Fig:cond_probab_sign}) and 
remains stable increasing time. \\ 
The above results are consistent with the empirical observations reported by 
Bouchaud et al~\cite{Bouchaud}. They consider a set of 450 U.S. equities, 
obtaining the same shape of fig.~\ref{Fig:cond_probab_sign}, with a 90\% 
of the stocks have the same sign as that of the market return for very large 
market movements.  \\
Figure~\ref{Fig:corr} clearly shows that correlations between stocks are 
enforced during high-volatility periods. 
Indeed in a one step model for an ensemble of $N$ stocks, it is always true 
that the mean stocks cross-correlation, $\rho$, is related to market 
variance, $\sigma_M^2=\langle M^2 \rangle - \langle M \rangle^2$ (over a 
horizon of one step), through the relation: 
$\rho \approx \sigma_M^2 - \frac{1}{N}$. However in our model the spreading 
of market volatility is enormously accentuated respect to the uncorrelated 
case (see fig.~\ref{Fig:corr} for a comparison). 
As a consequence, in our model, we have effectively periods characterized by 
high volatility and strong cross-correlations alternated to normal market 
activities. 
%
%
%
%----> Fig. 4
%
%
%----> End fig. 4

%-----------------> Section: Summary and Conclusions
%
%
\section{Summary and Conclusions} \label{section:conclusions}

In this paper we have proposed a new model that aims to capture some of most 
puzzling aspects of statistical properties of financial time series. 
Specifically, we have focalised our attention on two open problems: 
(i) leptokurtic behaviours in the probability distribution of stock returns and 
(ii) inter-dependence between market volatility and correlation structure 
when an ensemble of equities is considered. 
Both aspects cannot be taken into account by the standard approach commonly 
used in finance to describe correlated risk factors (that is the covariate 
Gaussian model). \\
We have introduced a new model that encompasses in a coherent picture both 
points (i) and (ii). The mechanism we have proposed is based on the interplay 
between equities movements and market trend. 

The results obtained suggest two considerations: (i) fat tails and correlation 
structure of extreme returns could be indeed strictly connected; (ii) an 
ensemble of stocks can be regarded as a ``critical system'', where the 
collective inter-dependences among equities play a key role.  

Our analysis may be of interest, from a theoretical point of view, in 
clarifying statistical properties of ensembles of stocks and, from an
applied perspective, in improving portfolio market risk estimations. 
% in Risk Management (e.g. to improve market risk estimations
% and make a more rational price evaluation of derivative products) 

%-----------------> Section: appendix 
%
%
\appendix 
\section{} \label{section:appendix}

This short appendix is devoted to determine the long time scaling behaviour of 
the solution of PDE~(\ref{eq:PDE_h}), satisfying the initial condition: 
$h(M,t=0) = 0$. \\
For $1/N = 0$ a close solution of equation~(\ref{eq:PDE_h}) can be worked out:
\begin{equation}
h^{(0)}(M,t) = \sqrt{ \frac{2}{\gamma} } \, \frac{M}{\mid M \mid} \left( 
\sqrt{t + \frac{1}{2 \gamma M^2} } - \sqrt{\frac{1}{2 \gamma M^2} } \right) \;. 
\label{eq:h_inf} 
\end{equation}
On the other hand, although for finite $N$ an exact solution cannot be found, 
it is still possible to investigate the behaviour of $h$ in the long time 
regime. This is done by considering a perturbative solution of 
equation~(\ref{eq:PDE_h}) respect to parameter $\frac{1}{N}$: 
$h(M,t) = \sum_j \left( \frac{1}{N} \right)^j h^{(j)}(M,t)$. 
It is a matter of algebra to show that each function $h^{(j)}$ 
contains only terms of the form
$\left[ t +1/ \left( 2 \gamma M^2 \right)  \right]^{\frac{1}{2}-k}$, 
$0 \le k \le 2 j$; this suggests to group together equal terms and look for 
a solution of the form: 
\begin{eqnarray}
&& h(M,t) \, = \, \frac{M}{\mid M \mid} \, \sum_{k=0}^{\infty} f_{k}(M) 
\cdot \nonumber \\
\cdot && \left[ \left( t + \frac{1}{2 \gamma M^2} \right)^{\frac{1}{2}-k} - 
\left(\frac{1}{2 \gamma M^2} \right)^{\frac{1}{2}-k} \right] \;. 
\label{eq:h_sol_series_exp} 
\end{eqnarray}
It turns out that each function $f_{k}$, which indeed depends on $N$, 
satisfies the recursive equation:
\begin{eqnarray}
&-&2 \, \gamma \, N \, \frac{M^3}{1-M^2} \, \frac{d f_{k}}{dM} + 
\frac{d^2 f_{k}}{dM^2} + \nonumber \\
&+& \left( \frac{3}{2} -k \right) \, \left[ \frac{3}{\gamma M^4} \, f_{k-1} 
- \frac{2}{\gamma M^3} \, \frac{d f_{k-1}}{dM} \right] + \nonumber \\
&+& \left( \frac{3}{2} -k \right) \, \left( \frac{5}{2} -k \right) \, 
\frac{1}{\gamma^2 M^6} \, f_{k-2} = 0 \;, 
\label{eq:recursive_for_f_k}
\end{eqnarray}
supplemented with the global condition:
\begin{equation}
\sum_{k=0}^{\infty} \left( 2\, \gamma \right)^{ k+\frac{1}{2} } \,  
\left( \frac{1}{2} -k \right) \, M^{2k} \, f_{k}(M) = 1 \;.
\label{eq:condition_on_f_k}
\end{equation}
It is possible to demonstrate the following statements about $\{f_k\}$: 
\begin{itemize}
	\item[(i)]  
	$f_0$ must be a constant;  
	\item[(ii)]  
	for any given $M$ and $k\geq 1\;$: 
	$\lim_{N \rightarrow \infty} f_{k}(M) = 0$;
	\item[(iii)]  
	for any given $M$: 
	$\lim_{N \rightarrow \infty} f_{2k+1}(M)/f_{2k}(M) = 0$; 
	\item[(iv)]  
	each $f_{k}$ is a regular function apart from a divergence in 
	$M=0$: $f_{k} \sim M^{-2k}$ ($k \geq 1$). (Hence for $M \rightarrow 0$ 
	all the terms in eq.~(\ref{eq:h_sol_series_exp}) behave like $1/M$.)  
\end{itemize}
Basing on the above properties, it is straightforward to argue that the long 
time scaling behaviour of the solution of PDE~(\ref{eq:PDE_h}) is given by the 
dominant term $f_0$ (i.e. the term corresponding to $N = \infty$ apart from a 
multiplicative coefficient), even for small values of $M$. 
In the language of Renormalization Group (RG), we say that the term in 
eq.~(\ref{eq:PDE_h}) proportional to $\frac{1}{N}$ is irrelevant. (An overview 
of the application of RG ideas to PDEs is given in~\cite{Bricmont}.)  \\
To summarise, for large times compare to $ \frac{1}{\gamma M^2}$, the solution 
of PDE~(\ref{eq:PDE_h}) behaves like $h^{(0)}(M,t)$. Hence for 
$t \gg \frac{1}{\gamma M^2}$, we have that $h(M,t) \sim t^{\frac{1}{2}}$ and a 
power law behaviour emerges in the model.

%-----------------> Acknowledgments 
%
%
% \acknowledgments

%-----------------> References
%
%

\end{document}